\documentclass[conference]{IEEEtran}
\usepackage{epsfig,latexsym,amsmath,amssymb,setspace,cite,color,graphicx,algorithm,algorithmic,cases}
\usepackage[caption=false,font=footnotesize]{subfig}
\usepackage[percent]{overpic}

\usepackage{physics}

\usepackage{amsmath}
\usepackage{amsfonts}
\usepackage{amssymb}
\usepackage{dsfont}
\usepackage{bm}
\usepackage{amsthm}
\usepackage{newlfont}
\usepackage{float}
\usepackage{hyperref}
\usepackage{algorithm}
\usepackage{algorithmic}
\usepackage{enumerate}
\usepackage{chngcntr}
\usepackage{mathtools}
\usepackage{breqn}
\usepackage{stmaryrd}
\usepackage{cite}
\hypersetup{
    colorlinks=true, 
    linkcolor= blue, 
    citecolor=blue 
}

\begin{document}
\sloppy
\allowdisplaybreaks[1]


\newtheorem{thm}{Theorem} 
\newtheorem{lem}{Lemma}
\newtheorem{prop}{Proposition}
\newtheorem{cor}{Corollary}
\newtheorem{defn}{Definition}
\newcommand{\remarkend}{\IEEEQEDopen}
\newtheorem{remark}{Remark}
\newtheorem{rem}{Remark}
\newtheorem{ex}{Example}
\newtheorem{pro}{Property}

\newenvironment{example}[1][Example]{\begin{trivlist}
\item[\hskip \labelsep {\bfseries #1}]}{\end{trivlist}}

\renewcommand{\qedsymbol}{ \begin{tiny}$\blacksquare$ \end{tiny} }

\renewcommand{\algorithmicrequire}{\textbf{Input:}}
\renewcommand{\algorithmicensure}{\textbf{Inputs:}}

\renewcommand{\leq}{\leqslant}
\renewcommand{\geq}{\geqslant}


\title {Secure Distributed Storage: Rate-Privacy Trade-Off and XOR-Based Coding Scheme
	\thanks{This work was supported in part by the NSF under grants CCF-1850227 and CNS-1526547.}
}

\IEEEoverridecommandlockouts

\author{\IEEEauthorblockN{R\'emi A. Chou}
\IEEEauthorblockA{Department of Electrical Engineering \& Computer Science\\
Wichita State University\\Wichita, KS 67260\\remi.chou@wichita.edu
}
 \and 
 \IEEEauthorblockN{J\"org Kliewer}
 \IEEEauthorblockA{Department of Electrical \& Computer Engineering\\
New Jersey Institute of Technology\\Newark, NJ 07102\\jkliewer@njit.edu}
}

\maketitle
\begin{abstract}
We consider the problem of storing data in a distributed manner over $T$ servers. We require the data (i) to be recoverable from the $T$ servers, and (ii) to remain private from any $T-1$ colluding servers, where privacy is quantified in terms of mutual information between the data and all the information available at the $T-1$ colluding servers. For this model, we determine (i) the fundamental trade-off between storage size and the level of desired privacy, (ii) the optimal amount of local randomness necessary at the encoder, and (iii)~an explicit low-complexity coding scheme that solely relies on XOR operations and that asymptotically (with the data size) matches the fundamental limits found.
\end{abstract} 

\section{Introduction}
Secure distributed storage schemes, e.g.,~\cite{garay2000secure,rawat2016centralized,bitar2018staircase,yu2018lagrange,huang2016communication}, often rely on the idea of secret sharing as introduced in~\cite{shamir1979share,blakley1979safeguarding}. Hence, there is a fundamental lower bound on the required storage space necessary to securely store information in a distributed manner. Specifically, in any secret sharing scheme, the total amount of information that needs to be stored must at least be equal to the entropy of the secret times the number of participants, see e.g., \cite{beimel2011secret}, and it is thus impossible to reduce the storage space without any  changes to the model assumptions.

  To this end, we propose to determine the optimal cost reduction, in terms of storage space, that can be obtained in exchange of tolerating a \emph{controlled amount} of reduced privacy.  This idea is closely related to non-perfect secret sharing \cite{farras2016recent,yoshida2018optimal} with a non-linear access function that generalize traditional ramp secret sharing \cite{yamamoto1986secret,yoshida2007secure,blakley1984security}. Unfortunately, for large secrets, as required for data storage, no low-complexity coding scheme is known to implement non-perfect secret sharing. Note that in the absence of a privacy constraint, for the related problems of perfect secret sharing and secure distributed storage,  XOR-based coding schemes have been proposed in~\cite{wang2014efficient} and \cite{huang2016secure,huang2017secure}, respectively.

We aim to fill this void in this paper and focus on a setting where a file $F$ needs to be stored at $T$ servers. The data is intended to  be recoverable from these $T$ servers, and needs to remain private from any $T-1$ colluding servers. Here, privacy is quantified in terms of mutual information between the data and all the information available at the $T-1$ colluding servers. 

To be concrete, consider the example of three servers, i.e., $T=3$, where the mutual information between the data of any two servers and the file $F$ must not exceed $L \triangleq \frac{1}{4}H(F)$. Assuming that $F$ is a sequence of uniformly distributed bits, we split $F$ in four parts $(F_1,F_2,F_3,F_4)$ of equal length (for simplicity we assume here that $|F|$ is a multiple of four), and we store in the three servers the shares 
\begin{align*}
	M_1 & \triangleq (F_1  \lVert K_1 \lVert F_2 \oplus K_2 \oplus K_4 ),\\
	M_2 & \triangleq (K_2  \lVert K_3 \lVert F_3  \oplus K_1 \oplus K_5 ), \\
	M_3 & \triangleq (K_4  \lVert K_5 \lVert F_4 \oplus K_3),
\end{align*}
where $(K_1,K_2,K_3,K_4,K_5)$ are five sequences of uniformly distributed bits with size $|F|/4$, $\oplus$ denotes the XOR operation, and $\Vert$ denotes concatenation.
We remark that all the four parts $(F_1,F_2,F_3,F_4)$ are either stored in clear or encrypted through a one-time pad. By inspection, one easily sees in this example that $F$ can be recovered from $(M_1,M_2,M_3)$ and any two shares leak at most $\frac{1}{4}H(F)$ bits about $F$. As it will be shown in the following, the size of the shares is optimal as well as the amount of local randomness, i.e., the length of $(K_1,K_2,K_3,K_4,K_5)$.

Our main contribution is the design of a low-complexity coding scheme for this problem with arbitrary parameters $L$ and $T$ that solely relies on XOR operations and that is asymptotically (with file size) optimal in terms of data storage and the required amount of local randomness at the encoder.

The remainder of the paper is organized as follows. We formalize the problem in Section \ref{secs} and state our main results in Section \ref{secres}. Our coding scheme is presented in Section \ref{sec:CS}. We present the proofs  of our results in the appendix. Finally, we provide concluding remarks in Section \ref{secconcl}.


\section{Problem statement} \label{secs}

Notation: For $a,b \in \mathbb{R}$, define $\llbracket a,b \rrbracket \triangleq [\lfloor a \rfloor , \lceil b \rceil ] \cap \mathbb{N}$ and $[a]^+\triangleq\max(0,a)$. Let $\oplus$ denote the XOR operator.\\
Consider $T \geq 2$ servers and define $\mathcal{T} \triangleq \llbracket 1,T\rrbracket$. Consider a file $F$ which is a sequence of $|F|$ bits uniformly distributed over $\{0,1\}^{|F|}$.
\begin{defn}
A $(\lambda,\rho)$ coding scheme consists of 
\begin{itemize}
		\item A stochastic encoder $e : \{ 0,1\}^{|F|} \times \{ 0,1\}^{\rho} \to \{ 0,1\}^{\lambda T}, (F,R) \mapsto  (M_t)_{t \in \mathcal{T}}$, which takes as input the file $F$ to store and a sequence $R$ of $\rho \in \mathbb{N}$ bits uniformly distributed over $\{ 0,1\}^{\rho}$ and independent of $F$, and outputs $T$ sequences (referred to as shares in the following) $(M_t)_{t \in \mathcal{T}}$ of length $\lambda \in \mathbb{N}$, where $M_t$ is stored in Server $t \in \mathcal{T}$.
	\item A decoder $d : \{ 0,1\}^{\lambda T} \to  \{ 0,1\}^{|F|}, (M_t)_{t \in \mathcal{T}} \mapsto \widehat{F}$, which takes as input all the $T$ sequences stored at the servers, and outputs an estimate $\widehat{F}$ of the file $F$.
\end{itemize}	
\end{defn}
\begin{defn}
For $\alpha \in [0,1]$, a $(\lambda,\rho)$ coding scheme is said to be $\alpha$-private if
\begin{align}
	H(F|M_{\mathcal{T}}) & = 0, \text{ (Decodability)} \label{eqreq1} \\
	\lim_{|F| \to \infty } \frac{I(F;M_{\mathcal{S}})}{H(F)} &\leq \alpha, \forall \mathcal{S} \subsetneq \mathcal{T}, \text{ (Privacy)} \label{eqreq2}
\end{align}
where we have used the notation $M_{\mathcal{S}} \triangleq (M_t)_{t \in \mathcal{S}}$, $\forall \mathcal{S} \subsetneq \mathcal{T}$.
\end{defn}
The objective is to design $\alpha$-private $(\lambda,\rho)$ coding schemes with minimal storage size requirement, i.e., minimal $\lambda$, and minimal amount of local randomness requirement at the encoder, i.e., minimal $\rho$.

In the following, one can assume 
	\begin{align} \label{assumption1}
		\alpha < 1 - 1/T.
	\end{align}
 Indeed, if $\alpha \geq 1- 1/T$, i.e., $\alpha \geq \frac{T-1}{T}$, then the privacy constraint is trivially satisfied if one splits the file in $T$ parts of size  $H(F)/T$ and store one part in each server.

\section{Main results} \label{secres}
\begin{thm}[Minimal storage size requirement] \label{thconverse}
  For any $\alpha$-private $(\lambda,\rho)$ coding scheme, we have
\begin{align*}
 \lim_{|F| \to \infty}	\frac{\lambda}{H(F)} \geq  (1-\alpha).
\end{align*}	
\end{thm}

\begin{thm}[Minimal local randomness requirement] \label{thconverse2}
For any $\alpha$-private $(\lambda,\rho)$ coding scheme, we have
\begin{align*}
 \lim_{|F| \to \infty}	\frac{\rho}{H(F)} \geq [T(1-\alpha) -1]^+.
\end{align*}	
\end{thm}

\begin{thm}[Achievability] \label{th_achiev}
Assume that $\alpha = 0$ or $\alpha = \frac{l}{k} $, for some $k,l\in\mathbb{N}^*$ with $k$ and $l$ coprime, and $\alpha \in [0,1-1/T[$. By density of $\mathbb{Q}$ in $\mathbb{R}$, such an $\alpha$ can be chosen arbitrarily close to any point in $[0,1]$.
The coding scheme in Section \ref{sec:CS} is an $\alpha$-private $(\lambda,\rho)$ coding scheme with
\begin{align}
\lim_{|F| \to \infty}	\frac{\lambda}{H(F)} &=  (1-\alpha), \label{eqL}\\
 \lim_{|F| \to \infty}	\frac{\rho}{H(F)}& = [T(1-\alpha) -1]^+.
 \label{eqR}
\end{align}	
\end{thm}

\section{Coding scheme} \label{sec:CS}
Consider $\alpha \in [0,1-1/T[$ such that $\alpha =0$ or  $\alpha = \frac{l}{k}$, for some $k,l\in\mathbb{N}^*$ with $k$ and $l$ coprime.

\subsection{Preliminaries}
\begin{itemize}
\item There exists $q \in \mathbb{N}$, $r \in \llbracket 0,T-2 \rrbracket$ such that 
	\begin{align} \label{div1}
		l = q(T-1) + r,
	\end{align}
and one can assume
	\begin{align} \label{assumption2}
		qT+r < k.
	\end{align}   
	Otherwise, if $qT+r = k$ (one has a similar argument if $qT+r > k$), then one can split the file $F$ in $k$ parts of equal size and store $q+1$ parts of $F$ in the first $r$ servers and $q$ parts of $F$ in the remaining servers. The privacy constraint is satisfied by \eqref{div1} because any $T-1$ servers have at most $q(T-1)+r=l$ parts of $F$, and \eqref{eqreq2} holds.
	\item  Note that $T(k-l)-k =(T-1)k -lT>  0$ by \eqref{assumption1}, and there exists $u \in \mathbb{N}$, $v \in \llbracket 0,T-1 \rrbracket$ such that 
	\begin{align} \label{div2}
		T(k-l)-k = u T + v.
	\end{align}
\end{itemize}
We emphasize that $l$, $k$, and $T$ are the only parameters of the coding scheme; $q$, $r$, $u$, and $v$ are obtained as sole functions of $l$, $k$, and~$T$.

\subsection{Coding Scheme}
\noindent{}\textbf{Step 1}. Divide the file in $k$ parts $(F_i)_{i \in \llbracket 1 , k \rrbracket}$ (if necessary, add $\beta$ zeros to $F$, where $\beta$ is the smallest integer in $\llbracket 1 , k-1 \rrbracket$ such that $(H(F)+ \beta)/k \in \mathbb{N}$. For convenience, we write $F_{i:j} \triangleq (F_{i'})_{i' \in \llbracket i , j \rrbracket}$ for $i,j \in \llbracket 1 , k\rrbracket$.\\ 
\textbf{Step 2}. Generate $N_{\textup{keys}} \triangleq T(k-l)-k$ ($>0$ by~\eqref{div2}) keys $(K_i)_{i \in \llbracket 1, N_{\textup{keys}} \rrbracket}$ each uniformly distributed over $\{0,1\}^{(H(F)+ \beta)/k}$ and independent of all other random variables.  For convenience, we write $K_{i:j} \triangleq (K_{i'})_{i' \in \llbracket i , j \rrbracket}$ for $i,j \in \llbracket 1 , N_{\textup{keys}}\rrbracket$. \\
\textbf{Step 3}. We now describe how to design the shares $M_{\mathcal{T}}$. 
Each share $M_t$, stored in Server $t\in \mathcal{T}$, is a vector of $(k-l)$ sequences (labeled from $1$ to $k-l$) of size $(H(F)+ \beta)/k$.  
For convenience, for $t\in \mathcal{T}$, and $i,j \in \llbracket 1 , k-l\rrbracket$ such that $i\leq j$,  we write $M_t[i:j]$ to designate the sequences of $M_t$ labeled from $i$ to $j$, and $M_t[i]$ to designate the sequence of $M_t$ labeled by $i$.
For $t \in \mathcal{T}$, the $k-l$ components of the vector $M_t$ are of one of the following types. 
\begin{itemize}
\item An unencrypted part of $F$, i.e., an element of $\{F_i :i \in \llbracket 1 , k \rrbracket \}$. 
\item A key, i.e.,  an element of $\{ K_i : i \in \llbracket 1, N_{\textup{keys}} \rrbracket \}$;
\item An encrypted part of $F$, obtained by XORing a part of $F$ with one or several keys.
\end{itemize}

To precisely describe how the shares $M_{\mathcal{T}}$ are chosen we distinguish two cases.

\textbf{Case 1}. Assume that $r+v < T$.
The unencrypted parts of $F$ and keys are assigned according to Algorithms \ref{alg1}, \ref{alg2}, respectively. The encrypted parts of $F$ are defined and assigned according to Algorithm \ref{alg3}. Note that we have the following.
\begin{itemize}
	\item For $t \in \llbracket 1, r \rrbracket$, $M_t$ consists of $k-l$ sequences: $q+1$ unencrypted parts of $F$, $u$ keys, and \begin{align} \label{eqx}
	x \triangleq k-l -q -1-u
	\end{align}   encrypted parts of $F$.
	\item For $t \in \llbracket r+1, T-v \rrbracket$, $M_t$ consists of $k-l$ sequences: $q$ unencrypted parts of $F$, $u$ keys, and $x+1$ encrypted parts of $F$.
	\item For $t \in \llbracket T-v +1, T \rrbracket$, $M_t$ consists of $k-l$ sequences: $q$ unencrypted parts of $F$, $u+1$ keys, and $x$ encrypted parts of $F$.
\end{itemize}

 \textbf{Case 2}. Assume that $r+v \geq T$.
The unencrypted parts of $F$ and keys are assigned according to Algorithms \ref{alg1}, \ref{alg2b}, respectively. The encrypted parts of $F$ are defined and assigned according to Algorithm \ref{alg3b}. Note that we have the following.
\begin{itemize}
	\item For $t \in \llbracket 1, T-v \rrbracket$, $M_t$ consists of $k-l$ sequences: $q+1$ unencrypted parts of $F$, $u$ keys, and $x$ encrypted parts of~$F$.
	\item For $t \in \llbracket T-v+1, r \rrbracket$, $M_t$ consists of $k-l$ sequences: $q+1$ unencrypted parts of $F$, $u+1$ keys, and $x-1$ encrypted parts of $F$.
	\item For $t \in \llbracket r+1,T \rrbracket$, $M_t$ consists of $k-l$ sequences: $q$ unencrypted parts of $F$, $u+1$ keys, and $x$ encrypted parts of $F$.
\end{itemize}

\begin{rem}
In both cases, the first $N_{\textup{Plain}} \triangleq qT + r$ parts $(F_i)_{i \in \llbracket 1 , qT + r \rrbracket}$ are stored unencrypted in the servers, and the $N_{\textup{Encrypted}} \triangleq k - r - Tq$ ($>0$ by \eqref{assumption2}) remaining parts $(F_i)_{i \in \llbracket  qT + r + 1, k \rrbracket}$ are first encrypted using the keys $(K_i)_{i \in \llbracket 1, T(k-l)-k \rrbracket}$ before being stored in the servers. 
\end{rem}
\begin{rem}
In Case 1, we have $x\geq 0$, otherwise $k-r-Tq =N_{\textup{Encrypted}} = (r+v)x + (T-r-v)(x+1)<0$, which contradicts~\eqref{assumption2}.	In Case 2, we also have have $x\geq 0$, otherwise $k-r-Tq =N_{\textup{Encrypted}} = (T-v+T-r)x + (r -T +v)(x-1)<0$, which again  contradicts~\eqref{assumption2}.
\end{rem}

\begin{algorithm}
  \caption{Assignment of unencrypted parts}
  \label{alg1}
  \begin{algorithmic}[1] 
\REQUIRE File $F$
\FOR{$t \in \mathcal{T}$}  
   \STATE $M_t [1:q] \triangleq F_{(t-1)q+1:tq}$ 
\IF{$t \leq r$}  
   \STATE $M_t [q+1] \triangleq F_{Tq+t}$ 
\ENDIF
\ENDFOR
  \end{algorithmic}  
\end{algorithm}
\begin{algorithm}
  \caption{Assignment of keys when $r+v < T$}
  \label{alg2}
  \begin{algorithmic}[1] 
\ENSURE Keys $(K_i)_{i \in \llbracket 1, N_{\textup{keys}} \rrbracket}$
\FOR{$t \in \mathcal{T}$}  
\IF{$t \leq r$}  
   \STATE $M_t [q+2:q+1+u] \triangleq K_{u(t-1)+1:ut}$ 
\ELSIF{$r < t \leq T - v$}
	\STATE $M_t [q+1:q+u] \triangleq K_{u(t-1)+1:ut}$ 
\ELSIF{$  t > T - v $}
	\STATE $M_t [q+1:q+u] \triangleq K_{u(t-1)+1:ut}$
	\STATE $M_t [q+u+1] \triangleq K_{Tu + t - (T-v)}$ 
\ENDIF
 \STATE Define\\$\mathcal{I}_t \triangleq \{ j \in \llbracket 1,N_{\textup{keys}} \rrbracket : \exists i \in \llbracket 1, l-k \rrbracket, M_t[i]=K_j \}$
\ENDFOR
  \end{algorithmic}  
\end{algorithm}
\begin{algorithm}
  \caption{Assignment of keys when $r+v \geq T$}
  \label{alg2b}
  \begin{algorithmic}[1] 
\ENSURE Keys $(K_i)_{i \in \llbracket 1, N_{\textup{keys}} \rrbracket}$
\FOR{$t \in \mathcal{T}$}  
\IF{$t \leq T-v$}  
   \STATE $M_t [q+2:q+1+u] \triangleq K_{u(t-1)+1:ut}$ 
\ELSIF{$T-v < t \leq r$}
	\STATE $M_t [q+2:q+u+2]$\\ $\triangleq K_{u(T-v)+(u+1)(t- T +v -1)+1:u(T-v)+(u+1)(t- T +v)}$ 
\ELSIF{$ t > r $}
	\STATE $M_t [q+1:q+u+1]$ \\ $\triangleq K_{u(T-v)+ (u+1)(t-T+v-1) + 1:u(T-v)+ (u+1)(t-T+v)}$
\ENDIF
 \STATE Define\\$\mathcal{I}_t \triangleq \{ j \in \llbracket 1,N_{\textup{keys}} \rrbracket : \exists i \in \llbracket 1, l-k \rrbracket, M_t[i]=K_j \}$
\ENDFOR

  \end{algorithmic}  
\end{algorithm}

\begin{algorithm}
  \caption{Creation and assignment of encrypted parts when $r+v < T$}
  \label{alg3}
  \begin{algorithmic}[1] 
\ENSURE File $F$ and keys $(K_i)_{i \in \llbracket 1, N_{\textup{keys}} \rrbracket}$
\FOR{$i \in \llbracket 1, x +1 \rrbracket$}
\FOR{$t \in \mathcal{T}$}  
\IF{$t \leq r$}  
	\STATE $j \triangleq  N_{\textup{Plain}} + (t-1)x+ i$ 
	\STATE $z \triangleq  \begin{cases} q+u+1+i & \text{if }i \neq x+1 \\ \emptyset  & \text{if }i = x+1 \end{cases}$
\ELSIF{$r < t \leq T - v$}
	\STATE $j\triangleq N_{\textup{Plain}} + rx + (t-r-1) (x+1)+ i$
	\STATE $z\triangleq  q+u+i$
\ELSIF{$  t > T - v $}
	\STATE $j\triangleq N_{\textup{Plain}} + rx + (x+1)(T-v-r) + (t -T+v-1)x+ i$ 
		\STATE $z\triangleq  \begin{cases} q+u+1+i & \text{if }i \neq x+1 \\ \emptyset  & \text{if }i = x+1 \end{cases}$  
\ENDIF
\STATE For~$t'\in \mathcal{T} \backslash \{t\}$, choose the key $K_{a}$ with the smallest $a \in \mathcal{I}_{t'}$, which has not previously been chosen in this algorithm, among the keys stored in Server $t'$, and define $\kappa_{t'} \triangleq K_{a}$. If this is not possible, then define $\kappa_{t'} = \mathbf{0}$. 
\STATE If $z \neq \emptyset$, then define $M_t[z] \triangleq F_{j} \oplus \bigoplus_{t'\in \mathcal{T} \backslash \{t\}}  \kappa_{t'}$  
\ENDFOR
\ENDFOR
  \end{algorithmic}  
\end{algorithm}
\begin{algorithm}
  \caption{Creation and assignment of encrypted parts when $r+v \geq T$}
  \label{alg3b}
  \begin{algorithmic}[1] 
\REQUIRE File $F$
\FOR{$i \in \llbracket 1, x \rrbracket$}
\FOR{$t \in \mathcal{T}$}  
\IF{$t \leq T-v$}  
	\STATE $j\triangleq  N_{\textup{Plain}} + (t-1)x+ i$ 
	\STATE $z \triangleq   q+u+1+i$
\ELSIF{$T-v <t \leq r$}
	\STATE $j\triangleq N_{\textup{Plain}} + (T-v)x + (t-T+v-1) (x-1)+ i$
	\STATE $z\triangleq  \begin{cases}q+u+2+i & \text{if }i \neq x \\ \emptyset  & \text{if }i = x \end{cases}$  
\ELSIF{$  t > r$}
	\STATE $j\triangleq N_{\textup{Plain}} + (T-v)x + (r -T - v)(x-1) + (t -r-1)x+ i$ 
		\STATE $z\triangleq   q+u+1+i$  
\ENDIF
\IF{$z \neq \emptyset$}
\STATE For~$t'\in \mathcal{T} \backslash \{t\}$, choose the key $K_{a}$ with the smallest $a \in \mathcal{I}_{t'}$, which has not previously been chosen in this algorithm, and define $\kappa_{t'} \triangleq K_{a}$. If this is not possible, then define $\kappa_{t'} = \mathbf{0}$. 
\STATE Define $M_t[z] \triangleq F_{j} \oplus \bigoplus_{t'\in \mathcal{T} \backslash \{t\}}  \kappa_{t'}$  
\ENDIF
\ENDFOR
\ENDFOR
  \end{algorithmic}  
\end{algorithm}

\subsection{Examples}

\begin{ex} [$T=3$, $\alpha= \frac{3}{10}$]
Write $F=(F_i)_{i \in \llbracket 1 , 10 \rrbracket}$. We have $(l,k) \triangleq (3,10)$. By \eqref{div1}, we have $(q,r) = (1,1)$ and by~\eqref{div2}, we have $(u,v) = (3,2)$. We have $r+v =T$, so we are in Case 2.  Moreover, $N_{\textup{Keys}} = 11$, $N_{\textup{Plain}} = 4$, $N_{\textup{Encrypted}} = 6$, $x=2$.
After running Algorithms \ref{alg1} and \ref{alg2b}, we have
\begin{align*}
    M_1 &\triangleq (F_1 \lVert  F_4 \lVert  K_1 \lVert K_2    \lVert K_3   \lVert M_1[6]  \lVert M_1[7] ),\\
	M_2 &\triangleq (F_2 \lVert K_4  \lVert  K_5 \lVert  K_6   \lVert K_7   \lVert M_2[6]  \lVert M_2[7] ),\\
	M_3 &\triangleq (F_3 \lVert  K_8 \lVert K_9  \lVert K_{10} \lVert K_{11}\lVert M_3[6]  \lVert M_3[7]).
\end{align*}
After running Algorithm \ref{alg3b}, we obtain the encrypted parts of the file $F$ as
\begin{align*}
	M_1[6] &\triangleq  F_5 \oplus K_4 \oplus K_8, \\ 
	M_1[7] &\triangleq  F_6 \oplus K_6 \oplus K_{10} ,\\
	M_2[6] &\triangleq  F_7 \oplus K_1 \oplus K_9, \\
	M_2[7] &\triangleq  F_8 \oplus K_3 \oplus K_{11},\\
	M_3[6] &\triangleq  F_9 \oplus K_2 \oplus K_5,\\
	M_3[7] &\triangleq  F_{10} \oplus K_7 .
\end{align*}
\end{ex}

\begin{ex} [$T=5$, $\alpha = \frac{7}{11}$]
We have $(l,k) \triangleq (7,11)$. By \eqref{div1}, we have $(q,r) = (1,3)$ and by \eqref{div2}, we have $(u,v) = (1,4)$. We have $r+v > T$, so we are in Case 2.  Moreover, $N_{\textup{Keys}} = 9$, $N_{\textup{Plain}} = 8$, $N_{\textup{Encrypted}} = 3$, $x=1$.
After running Algorithms \ref{alg1} and \ref{alg2b} we have
\begin{align*}
	M_1 & \triangleq (F_1  \lVert F_6 \lVert K_1   \lVert M_1[4] ),\\
	M_2 & \triangleq (F_2  \lVert F_7 \lVert K_{2} \lVert K_{3} ), \\
	M_3 & \triangleq (F_3  \lVert F_8 \lVert K_4   \lVert K_{5}),\\
	M_4 & \triangleq (F_4  \lVert K_6 \lVert K_{7} \lVert M_4[4]), \\
	M_5 & \triangleq (F_5  \lVert K_8 \lVert K_9   \lVert M_5[4] ).
\end{align*}
After running Algorithm \ref{alg3b} the encrypted parts of $F$ are  
\begin{align*}
	M_1[4] &\triangleq  F_{9} \oplus K_2 \oplus K_4 \oplus K_6 \oplus K_8, \\ 
	M_4[4] &\triangleq   F_{10}\oplus K_1 \oplus K_3 \oplus K_{5} \oplus K_{9} ,\\
	M_5[4] &\triangleq  F_{11} \oplus K_{7}.
\end{align*}
\end{ex}

\begin{ex} [$T=3$, $\alpha=\frac{5}{17}$]
We have $(l,k) \triangleq (5,17)$. By \eqref{div1}, we have $(q,r) = (2,1)$ and by \eqref{div2}, we have $(u,v) = (6,1)$. We have $r+v < T$, so we are in Case~1.  Moreover, $N_{\textup{Keys}} = 19$, $N_{\textup{Plain}} = 7$, $N_{\textup{Encrypted}} = 10$, $x=3$. After running Algorithms~\ref{alg1} and \ref{alg2}, we have
\begin{align*}
M_1  \triangleq &(F_1 \lVert  F_2 \lVert  F_7  \lVert K_1  \lVert K_2  \lVert K_3 \lVert K_4  \lVert K_5 \lVert K_6 \\
	& \phantom{-----------}\lVert M_1[10] \lVert M_1[11] \lVert M_1[12]  ),\\
M_2  \triangleq  & (F_3 \lVert F_4 \lVert    K_7      \lVert  K_8  \lVert K_9 \lVert K_{10}  \lVert K_{11} \lVert K_{12} \\
&\phantom{------lll-}\lVert M_2[9] \lVert M_2[10]  \lVert M_2[11] \lVert M_2[12]  ),\\
M_3  \triangleq 	&(F_5  \lVert  F_6   \lVert K_{13}  \lVert K_{14} \lVert K_{15} \lVert K_{16}  \lVert K_{17} \lVert K_{18} \lVert K_{19} \\
& \phantom{-----------}\lVert M_3[10] \lVert M_3[11] \lVert M_3[12]).
\end{align*}
After running Algorithm \ref{alg3}, the encrypted parts of $F$ are obtained as follows:
\begin{align*}
	M_1[10] &\triangleq   F_8 \oplus K_7 \oplus K_{13},\\
	M_1[11] &\triangleq  F_9 \oplus K_9 \oplus K_{15},\\
	M_1[12] &\triangleq  F_{10} \oplus K_{11} \oplus K_{ 17}, \\
	M_2[9] &\triangleq  F_{11} \oplus K_1 \oplus K_{14}, \\
	M_2[10] &\triangleq  F_{12} \oplus K_3 \oplus K_{16}, \\
	M_2[11] &\triangleq  F_{13} \oplus K_{5} \oplus K_{18}, \\
	M_2[12] &\triangleq  F_{14} \oplus K_{19}, \\
	M_3[10] &\triangleq  F_{15} \oplus K_2 \oplus K_8, \\
	M_3[11] &\triangleq  F_{16} \oplus K_4 \oplus K_{10}, \\
	M_3[12] &\triangleq  F_{17} \oplus K_6 \oplus K_{12}.
\end{align*}
\end{ex}

\vspace{2ex}
\section{Concluding remarks} \label{secconcl}
We have studied the problem of storing a file in $T$ servers such that the privacy leakage generated by  $T-1$ colluding servers with respect to the content of the file is bounded.
The main contribution of this paper is a coding scheme for this problem that (i) achieves the asymptotically (with the file size) optimal storage space at the servers, (ii) uses the optimal amount of local randomness at the encoder, (iii) solely relies on XOR operations and is thus suited to handle large amount of data with low-complexity. Generalization of our XOR-based coding scheme to a threshold access structure, i.e., when decodability in \eqref{eqreq1} and the privacy constraint in \eqref{eqreq2} are replaced by \begin{align*}
	H(F|M_{\mathcal{A}}) & = 0, \forall \mathcal{A} \subset \mathcal{T} \text{ s.t. } |\mathcal{A}|\geq t,  \\
	\lim_{|F| \to \infty} \frac{I(F;M_{\mathcal{U}})}{H(F)} &\leq \alpha, \forall \mathcal{U} \subset \mathcal{T} \text{ s.t. } |\mathcal{U}| \leq t-1 \end{align*} for some $t \in \llbracket 1, T-1 \rrbracket$, is under investigation.

\appendices

\section{Proof of Theorem \ref{thconverse}}
\noindent{}Let $\mathcal{S} \subsetneq \mathcal{T}$. For any $\alpha$-private $(\lambda,\rho)$ coding scheme, we have
\begin{align*}
\lambda 
&\geq H(M_t) \displaybreak[0]\\ 
& \geq I(F;M_t|M_{\mathcal{T} \backslash \{ t\}}) \displaybreak[0]\\
       &  = H(F|M_{\mathcal{T} \backslash \{ t\}}) -  H(F|M_{\mathcal{T}}) \\
       & \stackrel{(a)}{=} H(F|M_{\mathcal{T} \backslash \{ t\}}) \\
       & = H(F) - I(F;M_{\mathcal{T} \backslash \{ t\}}) \\
       & \stackrel{(b)}{\geq}H(F) - H(F)(\alpha + o(1)) \\
       & =   (1-\alpha)H(F) + o(H(F)),
\end{align*}
where $(a)$ holds by \eqref{eqreq1}, $(b)$ holds by \eqref{eqreq2}. 

\section{Proof of Theorem \ref{thconverse2}}
For any $\alpha$-private $(\lambda,\rho)$ coding scheme, we have
\begin{align*}
 \rho + H(F)
& \stackrel{(a)}{=} H(R) + H(F)\\ 
& \stackrel{(b)}{=}  H(RF) \\
&\stackrel{(c)}{\geq}  H(M_{\mathcal{T}}) \\
& \stackrel{(d)}{=} \sum_{t\in\mathcal{T}} H(M_t | M_{1:t-1}) \\
& \stackrel{(e)}{\geq} \sum_{t\in\mathcal{T}} H(M_t | M_{\mathcal{T}\backslash \{ t\}}) \\
& \geq \sum_{t\in\mathcal{T}} I(M_t;F | M_{\mathcal{T}\backslash \{ t\}}) \\
& \stackrel{(f)}{\geq} \sum_{t\in\mathcal{T}} [(1-\alpha)H(F) + o(H(F))] \\
& = T (1-\alpha)H(F) + o(H(F)),
\end{align*}
where $(a)$ holds by uniformity of $R$, $(b)$ holds by independence between $F$ and $R$, $(c)$ holds because $M_{\mathcal{T}}$is a deterministic function of $(R,F)$, $(d)$ holds by the chain rule, $(e)$ holds because conditioning reduces entropy, $(f)$ holds by the proof of Theorem \ref{thconverse}. 

\section{Proof of Theorem \ref{th_achiev}} \label{seca}

We will use the following definition in our analysis of the coding scheme of Section \ref{sec:CS}.
\begin{defn}
Consider an encrypted part $E_t$ of $F$ stored in Server $t \in\mathcal{T}$ as in Line 14 of Algorithm \ref{alg3}, or Line 15 of Algorithm \ref{alg3b}. One can write $E_t$ as $ {E}_t = F_{j} \oplus \bigoplus_{t'\in \mathcal{T} \backslash \{t\}}  \kappa_{t'}$, where $j\in \llbracket 1, k \rrbracket$ and for $t'\in \mathcal{T} \backslash \{t\}$, $\kappa_{t'} \in \{K_i : i \in \mathcal{I}_{t'}\} \cup \{ \mathbf{0}\}$. The encrypted part $E_t$ is said to be protected by a key of Server $t' \in \mathcal{T}$, if there exists $i \in \mathcal{I}_{t'}$ such that $\kappa_{t'} = K_i$, i.e., if $\kappa_{t'} \neq \mathbf{0}$. 
\end{defn}
By construction, it is straightforward to verify that the storage size in \eqref{eqL} and the required amount of local randomness at the encoder in~\eqref{eqR} are satisfied. Next, we prove decodability~\eqref{eqreq1} and privacy \eqref{eqreq2}.

\emph{Decodability}: \eqref{eqreq1} holds because all the parts $(F_i)_{i \in \llbracket 1 , k \rrbracket}$ of $F$ are stored in $M_{\mathcal{T}}$: $N_{\textup{Plain}}$ parts are unencrypted and the $N_{\textup{Encrypted}}$ encrypted parts can be decrypted from modulo-2 addition with keys that are all stored in $M_{\mathcal{T}}$. 

\emph{Privacy}: It is sufficient to prove  that the privacy constraint in \eqref{eqreq2} holds  for all the subsets of $\mathcal{T}$ with size $T-1$, since $I(F;M_{\mathcal{S}'}) \leq I(F;M_{\mathcal{S}})$ if $\mathcal{S}' \subseteq \mathcal{S} \subseteq \mathcal{T}$.  We will thus prove that  \eqref{eqreq2} holds for the sets $\mathcal{S}_t \triangleq \mathcal{T} \backslash \{ t\}$, $t\in\mathcal{T}$. We first define $K_{\mathcal{S}_t}$, $F_{\mathcal{S}_t}$, and $E_{\mathcal{S}_t}$ as all the keys, unencrypted parts of $F$, and encrypted parts of $F$, respectively, stored in the servers whose indices are in $\mathcal{S}_t$. We next consider two cases.
\subsection{Case 1. $r+v < T$}
Remark that
\begin{align}
&(r-1)x + (T-v-r)(x+1) + vx  \nonumber\\ \nonumber
& = (T-1)x +T-v-r  \\ \nonumber
& \stackrel{(a)}{=}  (T-1)(k-l-q-1-u) + T-v-r \\
& \stackrel{(b)}{=} 1+u, \label{eqdif}
\end{align}
where $(a)$ holds by \eqref{eqx}, $(b)$ holds by \eqref{div1} and \eqref{div2}.	We next consider three cases depending on the value of $t \in \mathcal{T}$.
\begin{itemize}
	\item Assume $t \leq r$. We  have
	\begin{align}
&I(M_{\mathcal{S}_t}; F) \nonumber \\
& = I(K_{\mathcal{S}_t}F_{\mathcal{S}_t}E_{\mathcal{S}_t}; F) \nonumber \\ \nonumber
& \stackrel{(a)}{=} I(K_{\mathcal{S}_t}F_{\mathcal{S}_t} F_j (F_{s(i)} \oplus K_i )_{i \in \mathcal{I}_t}; F) \\ \nonumber
& = I(K_{\mathcal{S}_t}F_{\mathcal{S}_t} F_j; F) +  I((F_{s(i)} \oplus K_i )_{i \in \mathcal{I}_t}; F|K_{\mathcal{S}_t}F_{\mathcal{S}_t} F_j) \\ \nonumber
& \stackrel{(b)}{=} I(F_{\mathcal{S}_t} F_j; F) +  I((F_{s(i)} \oplus K_i )_{i \in \mathcal{I}_t}; F|K_{\mathcal{S}_t}F_{\mathcal{S}_t} F_j) \\ \nonumber
& \stackrel{(c)}{\leq} I(F_{\mathcal{S}_t} F_j; F) +  I((F_{s(i)} \oplus K_i )_{i \in \mathcal{I}_t}; (F_{s(i)} )_{i \in \mathcal{I}_t}) \\ \nonumber
& \stackrel{(d)}{\leq}  l (H(F)+ \beta)/k +  \textstyle\sum_{i \in \mathcal{I}_t} I(F_{s(i)} \oplus K_i ; F_{s(i)} ) \\
& \stackrel{(e)}{=} \alpha (H(F)+ \beta), \label{eqproof}
\end{align}	   
where $(a)$ holds  for some $j \in \llbracket N_{\textup{Plain}}+1,k \rrbracket$ and $(s(i))_{i \in \mathcal{I}_t} \in \llbracket N_{\textup{Plain}}+1,k \rrbracket^{|\mathcal{I}_t|} $ as follows. Note indeed that Server $t$ stores $|\mathcal{I}_t|=u$ keys and  all the other servers store $(r-1)x + (T-v-r)(x+1) + vx = u+1$ (by \eqref{eqdif}) encrypted parts of $F$, hence, considering the encrypted parts $E_{\mathcal{S}_t}$ of all the servers in $\mathcal{S}_t$, all \emph{but one} are protected by a key of Server~$t$ by Line 14 of Algorithm \ref{alg3}. $(b)$ holds because $K_{\mathcal{S}_t}$ is independent of $F$, $(c)$ holds by the chain rule and because $(K_{\mathcal{S}_t},(F_{i})_{i \in \llbracket 1 , k\rrbracket  \backslash \{ s(i) : i \in \mathcal{I}_t \}})$ is independent from $(F_{s(i)},F_{s(i)} \oplus K_i )_{i \in \mathcal{I}_t}$, $(d)$ holds because $F_{\mathcal{S}_t}$ contains $(r-1)(q+1) + (T-r)q = l-1$ (by~\eqref{div1}) parts of $F$, $(e)$ holds by the one-time pad \cite{stinson2005cryptography}.
	\item  Assume $ r+1 \leq t \leq T-v $. We have
		\begin{align*}
&I(M_{\mathcal{S}_t}; F) \\
& = I(K_{\mathcal{S}_t}F_{\mathcal{S}_t}E_{\mathcal{S}_t}; F) \displaybreak[0] \\
& \stackrel{(a)}{=} I(K_{\mathcal{S}_t}F_{\mathcal{S}_t} (F_{s(i)} \oplus K_i )_{i \in \mathcal{I}_t}; F) \\
& \stackrel{(b)}{\leq} \alpha (H(F)+ \beta), 
\end{align*}	
where $(a)$ holds  for some $(s(i))_{i \in \mathcal{I}_t} \in \llbracket N_{\textup{Plain}}+1,k \rrbracket^{|\mathcal{I}_t|} $ as follows. Note indeed that Server $t$ stores $|\mathcal{I}_t|=u$ keys, and all the other servers store $rx + (T-v-r-1)(x+1) + vx = u$ (by \eqref{eqdif}) encrypted parts of $F$. Hence, all the encrypted parts $E_{\mathcal{S}_t}$ of all the servers in $\mathcal{S}_t$ are protected by a key of Server~$t$ by Line 14 of Algorithm \ref{alg3}. $(b)$ holds similar to \eqref{eqproof} because $F_{\mathcal{S}_t}$ contains $r(q+1) + (T-r-1)q = l$ (by \eqref{div1}) parts of $F$.
	\item  Assume  $t \geq  T-v +1$. The proof of \eqref{eqproof} is identical to Subcase ii by remarking that (i) Server $t$ stores $|\mathcal{I}_t|=u+1$ keys, and all the other servers store $rx + (T-v-r)(x+1) + (v-1)x = u+1$ (by~\eqref{eqdif}) encrypted parts of $F$, and (ii) $F_{\mathcal{S}_t}$ contains $r(q+1) + (T-r-1)q = l$ (by \eqref{div1}) parts of $F$.
\end{itemize}
 
\subsection{Case 2. $r+v \geq T$}
Remark that
\begin{align}
&(T-v-1)x + (r-T+v)(x-1) + (T-r)x  \nonumber \\
&= (r-1)x + (T-v-r)(x+1) + vx  \nonumber \\
& =  1+u, \label{eqdif2}
\end{align}
where the last equality holds by \eqref{eqdif}. We next consider three cases depending on the value of $t \in \mathcal{T}$.
\begin{itemize}
	\item Assume $t \leq T-v$. We  have
	\begin{align}
&I(M_{\mathcal{S}_t}; F) \nonumber \\
& = I(K_{\mathcal{S}_t}F_{\mathcal{S}_t}E_{\mathcal{S}_t}; F) \nonumber \\ \nonumber
& \stackrel{(a)}{=} I(K_{\mathcal{S}_t}F_{\mathcal{S}_t} F_j (F_{s(i)} \oplus K_i )_{i \in \mathcal{I}_t}; F) \\ 
& \stackrel{(b)}{\leq} \alpha (H(F)+ \beta), 
 \label{eqproof2}
\end{align}	   
where $(a)$ holds  for some $j \in \llbracket N_{\textup{Plain}}+1,k \rrbracket$ and $(s(i))_{i \in \mathcal{I}_t} \in \llbracket N_{\textup{Plain}}+1,k \rrbracket^{|\mathcal{I}_t|} $ as follows. Note indeed that Server $t$ stores $|\mathcal{I}_t|=u$ keys and  all the other servers store $(T-v-1)x + (r-T+v)(x-1) + (T-r)x = u+1$ (by \eqref{eqdif2}) encrypted parts of $F$, hence, considering the encrypted parts $E_{\mathcal{S}_t}$ of all the servers in $\mathcal{S}_t$, all \emph{but one} are protected by a key of Server~$t$ by Line 15 of Algorithm~\ref{alg3b}. $(b)$ holds similar to \eqref{eqproof} because $F_{\mathcal{S}_t}$ contains $(r-1)(q+1) + (T-r)q = l-1$ (by \eqref{div1}) parts of $F$.
	\item  Assume $ r < t \leq T-v $. We have
		\begin{align*}
&I(M_{\mathcal{S}_t}; F) \\
& = I(K_{\mathcal{S}_t}F_{\mathcal{S}_t}E_{\mathcal{S}_t}; F) \\
& \stackrel{(a)}{=} I(K_{\mathcal{S}_t}F_{\mathcal{S}_t} F_j (F_{s(i)} \oplus K_i )_{i \in \mathcal{I}_t}; F) \\ 
& \stackrel{(b)}{\leq} \alpha (H(F)+ \beta), 
\end{align*}	
where $(a)$ holds  for some $j \in \llbracket N_{\textup{Plain}}+1,k \rrbracket$ and $(s(i))_{i \in \mathcal{I}_t} \in \llbracket N_{\textup{Plain}}+1,k \rrbracket^{|\mathcal{I}_t|} $ as follows. Note indeed that Server $t$ stores $|\mathcal{I}_t|=u+1$ keys, and all the other servers store $(T-v)x + (r-T+v-1)(x-1) + (T-r)x = u+2$ (by \eqref{eqdif2}) encrypted parts of $F$, hence, considering the encrypted parts $E_{\mathcal{S}_t}$ of all the servers in $\mathcal{S}_t$, all \emph{but one} are protected by a key of Server~$t$ by Line~15 of Algorithm~\ref{alg3b}. $(b)$ holds similar to \eqref{eqproof} because $F_{\mathcal{S}_t}$ contains $(r-1)(q+1) + (T-r)q = l-1$ (by \eqref{div1}) parts of $F$.
	\item  Assume  $t \geq  T-v +1$. We have
		\begin{align*}
&I(M_{\mathcal{S}_t}; F) \\
& = I(K_{\mathcal{S}_t}F_{\mathcal{S}_t}E_{\mathcal{S}_t}; F) \\
& \stackrel{(a)}{=} I(K_{\mathcal{S}_t}F_{\mathcal{S}_t} (F_{s(i)} \oplus K_i )_{i \in \mathcal{I}_t}; F) \\
& \stackrel{(b)}{\leq} \alpha (H(F)+ \beta), 
\end{align*}	
where $(a)$ holds  for some $(s(i))_{i \in \mathcal{I}_t} \in \llbracket N_{\textup{Plain}}+1,k \rrbracket^{|\mathcal{I}_t|} $ as follows. Note indeed that Server $t$ stores $|\mathcal{I}_t|=u+1$ keys, and all the other servers store $(T-v)x + (r-T+v)(x-1) + (T-r-1)x = 1+u$ (by \eqref{eqdif2}) encrypted parts of $F$. Hence, all the encrypted parts $E_{\mathcal{S}_t}$ of all the servers in $\mathcal{S}_t$ are protected by a key of Server~$t$ by Line~15 of Algorithm \ref{alg3b}. $(b)$ holds similar to \eqref{eqproof} because $F_{\mathcal{S}_t}$ contains $r(q+1) + (T-r-1)q = l$ (by \eqref{div1}) parts of $F$.
\end{itemize}

\bibliographystyle{IEEEtran}
\bibliography{bib}

\end{document}